\documentclass[%
reprint,
superscriptaddress,
 amsmath,amssymb,
 aps,
pra,
]{revtex4-2}
\usepackage[export]{adjustbox}
\usepackage{graphicx}
\usepackage{dcolumn}
\usepackage{bm}
\usepackage{amsfonts,amssymb,amsmath}
\usepackage[dvips]{epsfig}
\usepackage{hyperref}
\usepackage{upgreek}
\usepackage{mathrsfs}
\usepackage{xcolor}
\usepackage{bm} 
\usepackage{romannum}
\newcommand{\boldmathcal}[1]{\bm{\mathcal{#1}}} 

\begin{document}
\pagenumbering{arabic}
\preprint{APS/123-QED}

\title{Efficiency Enhancement up to Unity in a Generalized Quantum Otto Engine: Comparative Analysis with Conventional Quantum Otto Engine Utilizing a Two-Qubit Heisenberg XXZ Chain}

\author{Zorar Ahmadi}
 \email{zorarahmadi@gmail.com (Corresponding Author)}
\author{Bashir Mojaveri}
 \email{bmojaveri@azaruniv.ac.ir ; bmojaveri@gmail.com}
\affiliation{Department of Physics, Azarbaijan Shahid Madani University, PO Box 51745-406, Tabriz, Iran}

\date{\today}

\begin{abstract}

This study presents a comparative analysis of three quantum thermal engines utilizing a two-qubit Heisenberg XXZ chain as the working substance. A novel generalized quantum Otto cycle (GQOC) is introduced, featuring two distinct coupling configurations to thermal reservoirs. The GQOC exhibits the potential for 100\% efficiency, surpassing the efficiency of the conventional quantum Otto cycle. Essential conditions for positive work extraction and achieving maximum unity efficiency are derived. An experimental implementation using Quantum Electrodynamic circuits and Transmon qudit is proposed. This work contributes to the advancement of quantum heat engines, highlighting the benefits of non-equilibrium processes and asymmetric coupling for enhanced work extraction and efficiency.
\begin{description}
\item[Keywords]
Quantum Thermodynamics, Quantum Otto Engine, Generalized Quantum Otto Cycle
\end{description}
\end{abstract}

\maketitle

\section{\label{Intro}Introduction}

The concept of quantum heat engines (QHE) has attracted significant attention in recent decades due to remarkable advancements in quantum technologies \cite{Pet, Kla, Elo1, Rob, Jar, Yun, Tho, Hua, Kar, Must1, Hua2, Bar, Rei, Z4,Hat2024,Almeida24,chakmak23,ChakMus, Haddadi}. Traditionally, a classical heat engine is a cyclic device that converts heat into useful work \cite{Cal}. In the case of QHEs, the working fluid is a finite dimensional quantum system. By exploiting the principles of quantum mechanics, QHEs have the potential to exhibit unique characteristics and offer advantages over their classical counterpart. Some notable examples include the ability to extract work from a single heat bath \cite{Scu1}, surpassing the limits of the traditional Otto and Carnot cycles efficiencies \cite{Rob, Z4, Hua2, Scu1, Scu2, Klaers, Agr}, and optimizing efficiency by utilizing non-equilibrium reservoirs (NER) as a type of quantum resource \cite{Rob, Hua2, Klaers, Aba, San1, Nie, Assis}. Moreover, the proposal of novel quantum cycles can be implemented using quantum technologies, presents a paradigm for quantum machines with distinct quantum characteristics \cite{Hof, Shi, Das, Elo2}. 
Furthermore, QHEs contribute to the understanding of the significant connections between quantum correlations and the work output in a generalized quantum Otto cycle (GQOC) \cite{Hew}.\\

On the other hand, the presence of inherent asymmetry in the coupling between a system and its environment under non-equilibrium conditions has led to remarkable outcomes.These include the ability to block the flow of quantum thermal energy in quantum thermal diodes \cite{Ter, Per1, Per2, Bal1, Bal2, Meh, Nas, Z3, 3qbdYu} and the extraction of work through unitary cyclic processes in quantum batteries \cite{MojPRA}. In a recent study \cite{Z3}, we investigated a two-qubit system that is asymmetrically coupled to two distinct heat baths simultaneously, representing a perfect quantum thermal rectification with maximum entanglement. Building upon this work, our current study introduces a novel generalized version of the quantum Otto cycle, where a non-equilibrium interacting XXZ chain serves as the working substance. In contrast to conventional approaches that primarily focus on establishing quantum counterparts to classical thermodynamic cycles, the proposed cycle incorporates two adiabatic stages and two non-equilibrium steady-state stages, offering practical implementation possibilities. Specifically, we examine the impact of symmetric and asymmetric coupling between the system and the reservoirs during the non-equilibrium stages, and compare the performance of these configurations with that of a quantum thermal machine operating in a standard quantum Otto cycle.\\
In our investigation, we will demonstrate that in our specific GQOC, the extraction of positive work is solely contingent upon the geometric configuration of the system's coupling to the reservoirs. In essence, our findings indicate that our generalized quantum Otto machine can only operate and generate positive work when it is configured in an asymmetric manner. Moreover, we will illustrate that our GQOC exhibits the remarkable ability to achieve 100\% efficiency, which is not attainable in the conventional quantum Otto cycle (QOC). We will show that achieving 100\% efficiency in the GQOC does not violate the second law of thermodynamics. 
Furthermore, we derive a specific explicit expression that serves as an essential condition for extracting positive work in all three cycles. Moreover, through a comprehensive analytical discussion, we investigate the maximum achievable work output by considering the possible transition rates between the different energy levels of the system. In the case of the GQOC, we establish a specific analytical condition that enables the GQOC to achieve 100\% efficiency. Finally, we propose an experimental realization for our generalized quantum Otto machine using QED circuits and Transmon qudits, providing a practical framework for implementing and testing the proposed cycle.

\section{\label{cycles}Description of The Generalized Quantum Otto Cycle}
This section provides an explanation of the GQOC. A typical QOC involves four stages, including two adiabatic processes and two isochoric processes. However, in our specific GQOC, we replace the two isochoric processes with two distinct quantum processes (\emph{ non-equilibrium processes 1-2 and 3-4}). The performance of this cycle is described in detail below.
\begin{enumerate}
\item \emph{Non-equilibrium process 1-2}

The working substance, described by the Hamiltonian $H_S$ and a set of occupation probabilities $P_{i0}$ in each energy level, is brought into contact with two independent thermal reservoirs held at different temperatures. During this process, the system exchanges heat only with the hot and cold reservoirs, resulting in a change of the occupation probabilities from $P_{i0}$ to $P_i^{c}$ as a result of absorption or release of quantum heat from or into the reservoirs. The dynamics of the working substance can be described by the Markovian master equation \cite{Breuer, Lind, Scu-Zub}, within the Born-Markov approximation, assuming that the thermal reservoirs are in Gibbs states with finite temperatures $T_L$ and $T_R$. Eventually, the working substance reaches a steady state with occupation probabilities $P_i^{c}$ in each energy level.

\item \emph{Process 2-3 (Adiabatic process)}

In this stage, the system is isolated from the reservoirs and undergoes a quantum adiabatic expansion process where the degree of anisotropy changes from $\Delta^{c}$ to $\Delta^{h}$. The change must be sufficiently slow so that the thermal populations $P_i^{c}$ remain unchanged according to the quantum adiabatic theorem. As a result, the energy levels are varied from $E_i^{c}$ to $E_i^{h}$ and no heat is transferred, but work is done by the system or environment.

\item \emph{Non-equilibrium process 3-4}

The system, now with modified energy levels $E_i^{h}$ and occupation probabilities $P_i^{c}$, is reconnected to the same reservoirs, while the temperatures of the reservoirs are swapped (see (Fig.\ref{fig1}b and Fig.\ref{fig1}c). Consequently, the left part of the system, previously in contact with the hot reservoir, now links to the cold reservoir, and vice versa. During this stage, the energy levels remain unchanged, but the occupation probabilities change from $P_i^{c}$ to $P_i^{h}$, resulting in an exchange of heat between the system and the environment.

\item \emph{Process 4-1 (Adiabatic process)}

Finally, the working substance is disconnected from the reservoirs and its energy structure is changed back from $E_i^{h}$ to $E_i^{c}$ by altering $\Delta^{h}$ to $\Delta^{c}$ in an adiabatic contraction process. As a consequence of the adiabatic process, the probability of each eigenstate $P_i^{h}$ is maintained, and no heat is exchanged, but work is done. To complete the cycle, the condition $P_i^{h} = P_{i0}$ is required. According to the quantum mechanical interpretation of the first law of thermodynamics \cite{Kieu1, Kieu2}, the quantum heats transferred in the \emph{processes 1-2} and \emph{3-4} ($Q_{1-2}$, $Q_{3-4}$), net work done $W$, and operational efficiency $\eta$ can be expressed as: 

\begin{subequations}
\label{w:def}
\begin{eqnarray}
Q_{1-2}= \sum_i E_{i}^{c} \left[ P_{i}^{c}-P_{i}^{h} \right],\label{def:1}
\end{eqnarray}
\begin{eqnarray}
Q_{3-4}= \sum_i E_{i}^{h} \left[ P_{i}^{h}-P_{i}^{c} \right],\label{def:2}
\end{eqnarray}
\begin{eqnarray}
W= Q_{1-2}+Q_{3-4}=\sum_i \left(E_{i}^{h}-E_{i}^{c}\right)  \left[ P_{i}^{h}-P_{i}^{c} \right],\label{def:3}
\end{eqnarray}
\begin{eqnarray}
\eta=\frac{W}{Q_{in}}=\frac{W}{\sum Q_{>0}}.\label{def:4}
\end{eqnarray}
\end{subequations}
Here $\sum Q_{>0}$ refers to the total heat absorbed from the reservoirs and, $Q>0$ ($Q<0$) indicates the amount of quantum heat absorbed (released) from the reservoirs to the system (from the system to the reservoirs), and $W>0$ denotes the positive work performed by the machine. The condition $Q_{in}>(-Q_{out})>0$ is always hold for a quantum Otto cycle but not necessarily for our GQOC. As demonstrated at the end of Section \ref{sec 4}, it is possible for our GQOC to absorb energy during both non-equilibrium processes (1-2) and (3-4), resulting in $Q_{1-2}>0$ and $Q_{3-4}>0$.

\end{enumerate}

\subsection{\label{system}Working substance}

\begin{figure*}
\centering
\fbox{\includegraphics[trim={0 0 0 0},clip,width=\textwidth]{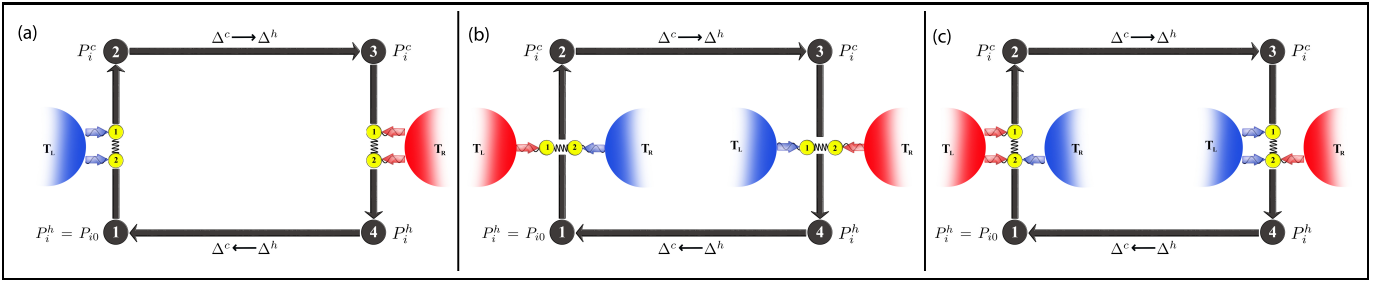}}
\caption{Schematic of three types of cycles. a) QOC. b) GQOC with symmetric coupling. c) GQOC with asymmetric coupling. }
\label{fig1} 
\end{figure*}

The main objective of this research is to introduce a new quantum cycle and compare it to the QOC, which is a nearly similar version. However, as the qubit-reservoir coupling configuration deeply affects its steady-state density matrix, it is necessary to study the GQOC in two different procedures. Fig.\ref{fig1} represents three different types of heat engines. In Fig.\ref{fig1}-a we depict the traditional quantum Otto engine. In processes (1-2 and 3-4) of this engine, the working substance is in contact with only one thermal reservoir. At the end of these processes, which are known as isochoric processes, the system attains thermal equilibrium with a bath of temperature $T_i$, and the population of its eigenstates is determined by the Boltzmann-Gibbs distribution. Processes 2-3 and 4-1 in the GQOC are similar to those in the QOC.\\
Fig.\ref{fig1}-b illustrates the second scenario, where the two-qubit system is linearly coupled with two different thermal baths. In contrast, the third case Fig.\ref{fig1}-c shows the system in contact with two baths in an asymmetric setting. Where, in the non-equilibrium process 1-2, qubit 1 interacts only with the hot reservoir (left reservoir) at temperature $T_L$, while qubit 2 is in contact with both the left and right reservoirs at temperatures $T_L$ and $T_R$, respectively. In the process 3-4, the temperatures of the reservoirs are changed, and qubit 1 is only in contact with the cold reservoir.
We exploit the two-qubit Heisenberg XXZ system as a working substance with the folowing Hamiltonian
\begin{eqnarray}
\label{Hs}
H_S=&&\frac{1}{2}[B\sigma_z^{(1)}+B\sigma_z^{(2)}+J\sigma_x^{(1)}\sigma_x^{(2)}+J\sigma_y^{(1)}\sigma_y^{(2)}\nonumber\\ 
&&+\Delta\sigma_z^{(2)}\sigma_z^{(2)}],
\end{eqnarray}
 where $B$ is the magnetic field in the direction of $z$ and $\sigma_x^{(i)}$, $\sigma_y^{(i)}$ and $\sigma_z^{(i)}$ are Pauli matrices associated with $i$th qubit. $J$ and $\Delta$ are real parameters that exhibit interqubit coupling and anisotropy of the system, respectively, \cite{Wan1, Orb}. The eigenstates and corresponding eigenenergies of the system are obtained as follows:
\begin{subequations}
\begin{eqnarray}
|\Phi_1\rangle=|0,0\rangle,\quad\quad\quad\quad\,\,\,\,\,\,\,\quad\quad\quad E_1=\frac{1}{2}\left(\Delta-2B\right),\label{energy1}\\
|\Phi_2\rangle=|1,1\rangle,\quad\quad\quad\quad\,\, \,\,\,\,\quad\quad\quad E_2=\frac{1}{2}\left(\Delta+2B\right),\label{energy2}\\
|\Phi_3\rangle=\frac{1}{\sqrt{2}}\left[|0,1\rangle-|1,0\rangle\right],\quad\,\, \,\,\,\,\,\,\,\,\,\quad E_3=-\frac{\Delta}{2}-J,\label{energy3}\\
|\Phi_4\rangle=\frac{1}{\sqrt{2}}\left[|0,1\rangle+|1,0\rangle\right],\quad\,\,\,\, \,\,\,\,\,\,\,\quad
E_4=-\frac{\Delta}{2}+J.\label{energy4}
\end{eqnarray}
\end{subequations}
Regarding the non-equilibrium processes in the above-mentioned cycle, it is essential to introduce the Hamiltonan of the thermal reservoirs in this way
\begin{eqnarray}
\label{HR}
H_R= \sum_j \omega_j^{(L)} a_j^{\dag} a_j+\omega_j^{(R)} b_j^{\dag} b_j,
\end{eqnarray}
where $a_j^{\dag}$ ($a_j$) and $b_j^{\dag}$ ($b_j$) are the creation (annihilation) operators of the $j$th bosonic mode with frequencies $\omega_j^{(L)}$ ($\omega_j^{(R)}$) of the left and right reservoirs, respectively. Here, the two reservoirs are assumed to be thermal and characterized by the inverse temperatures $\beta_i = \frac{1}{k T_{i}}, (i=L,R)$ where $k$ is the Boltzmann constant and $(L,R)$ refer to left and right respectively. Farthermore, disregarding the dephasing process, the symmetrically and asymmetrically dissipative interactions between the qubits and the reservoirs are explained by the Hamiltonian
\begin{eqnarray}
\label{HRS}
H_{R-S}=&&\sum_j \lambda_j^{(L)}\left( \sigma_x^{(1)}+\epsilon \sigma_x^{(2)} \right) \left(a_j^{\dag} + a_j \right)\\ \nonumber
&& + \lambda_j^{(R)}\sigma_x^{(2)}\left(b_j^{\dag} + b_j \right),
\end{eqnarray}
where the real parameter $\lambda_j^{(L)}$ ($\lambda_j^{(R)}$) shows the coupling strengths between the system and the left (right) thermal reservoir. We use here the parameter $\epsilon$ to represent the asymmetry in the system-reservoirs coupling. Where $\epsilon$=1 ($\epsilon$=0) refers to the asymmetric (symmetric) coupling. 
As mentioned before, the dynamics of such a non-equilibrium system could be described by the Markovian master equation. One can find the precise related steady-state solution of this system in Appendix (\ref{AppA}).\\

\section{\label{sec 3}Analyzing the cycles}
In the following we explore the conditions for positive work extraction in the QOC, GQOC with symmetric couplings, and GQOC with asymmetric couplings. To this end, we substitute the eigenenergies of the system in Eqs.(\ref{energy1}-\ref{energy4}) into Eq.(\ref{w:def}-c) and consider the varying of the parameter $\Delta$ during the adiabatic processes. This approach enables us to derive the necessary conditions for achieving positive work. For the case we use in this paper where $\Delta^{h} > \Delta^{c}$, it is straight forward to obtain the below condition
\begin{eqnarray}
\label{W condition}
\Xi_{3,4}=\sum_{i=3,4} \left(P_i^{c}-P_i^{h}\right) > \Xi_{1,2}=\sum_{j=1,2} \left( P_j^{c}-P_j^{h} \right).
\end{eqnarray}
This fundamental condition applies to all three cycles investigated in our study, revealing that positive work can only be extracted from each quantum engine if, the sum of the population differences between the cold and hot isochoric processes of the QOC (which correspond to two non-equilibrium processes in the GQOC) of two entangled states ($|\Phi_3\rangle$ and $|\Phi_4\rangle$) are larger than those of the unentangled states ($|\Phi_1\rangle$ and $|\Phi_2\rangle$). Notably, Eq.(\ref{W condition}) does not require that the populations of entangled states be higher than those of unentangled states for positive work extraction. Instead, it emphasizes the significance of maximizing the changes in the populations of entangled states relative to those of unentangled states, particularly during the non-equilibrium stages.
In order to determine which specific quantum engine can meet the aforementioned restriction, we investigate the performance of different engines under certain allowed parameter values, where the quantum Otto engine satisfies the second law of thermodynamics i.e. $Q_{in}>-Q_{out}>0$.\\

\begin{figure}
\centering \includegraphics[width=230pt,trim={2.2cm 0 4cm 0}]{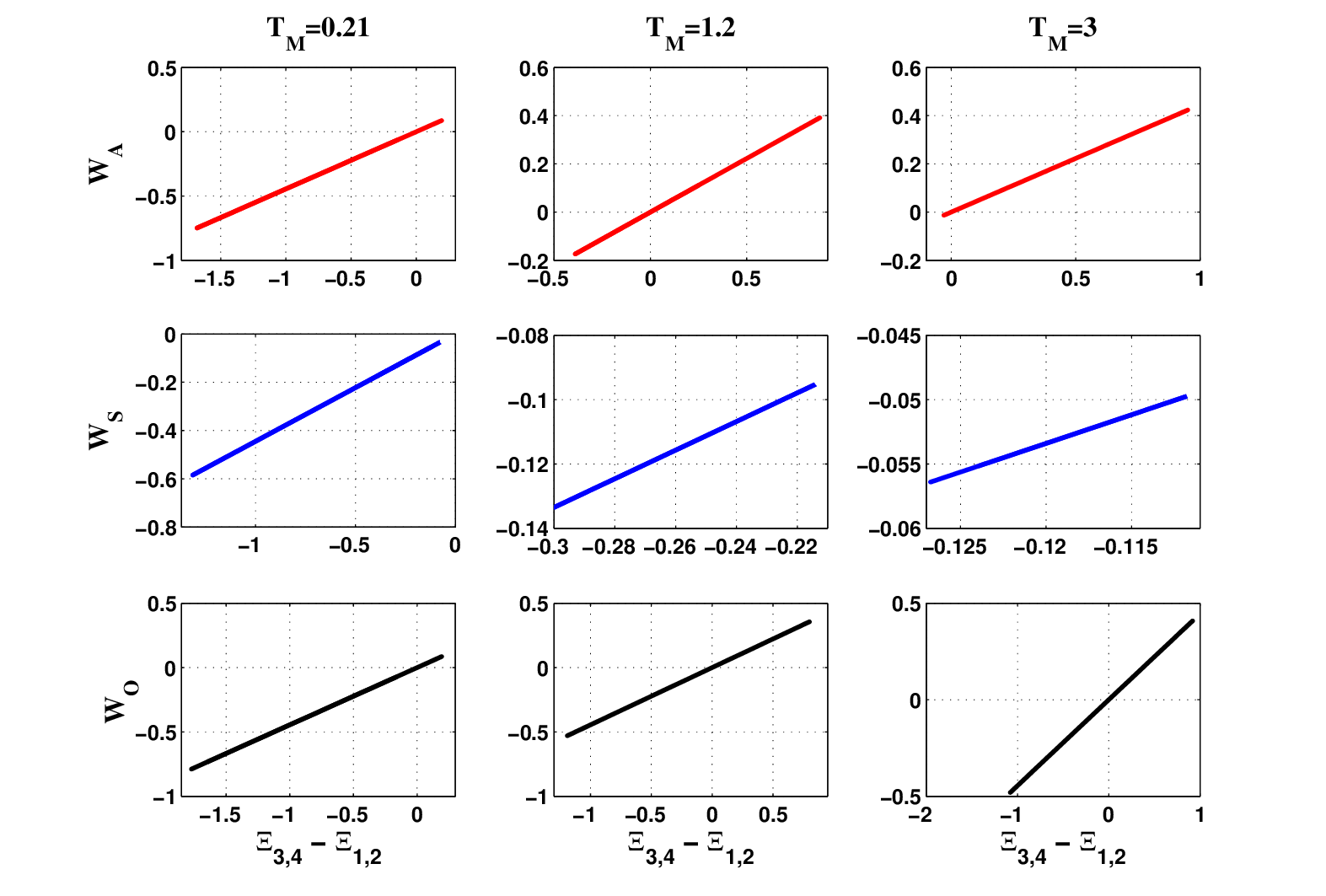}
\caption{The work output $W$ (in units of $\hbar J$) as a function of population differences for three different quantum engines at three different mean reservoir temperatures $T_M$. The work output of the GQOC with asymmetric coupling ($W_A$), GQOC with symmetric coupling ($W_S$), and the quantum Otto cycle ($W_O$) are presented in the first, second, and last rows, respectively, while each column corresponds to a different mean temperature. The model parameters are set to $\Delta^{c}=0.10$, $\Delta^{h}=0.99$, $\kappa=0.05$, $T_{cold}=0.005$ and $\Delta T=2T_M$, where all parameters are expressed in units of the interqubit coupling $J$.}
\label{fig2} 
\end{figure}

In Fig.\ref{fig2}, we present a plot of the work output $W$ (in units of $\hbar J$) as a function of the differences in population between the system eigenstates ($\Xi_{3,4}-\Xi_{1,2}$) for three different quantum cycles, while varying the magnetic field $B$ in the range of $\left[-3,3\right]$. The system parameters are fixed at $\Delta^{c}=0.10 J$ and $\Delta^{h}=0.99 J$, and natural units ($\hbar=k_B=1$) are used throughout the study. Additionally, we introduce the environmental parameters in a more convenient way by defining the mean temperature $T_{M}=\frac{T_L+T_R}{2}$ and the temperature gradient $\Delta T=T_L-T_R$ of the reservoirs. It is worth noting that, to avoid singularities and divergences, we set $T_{R}=T_{cold}=0.005$ throughout the paper.\\
The key insight gleaned from this figure is that, over the range of magnetic fields considered ($-3 \leq B \leq 3$), the GQOC with symmetric coupling cannot achieve $\Xi_{3,4} > \Xi_{1,2}$. As a result, positive work cannot be extracted from this cycle, rendering it unsuitable as a quantum heat engine. We note that, despite examining a wide range of permitted parameters, we consistently observed negative work output ($W_S$) for this cycle, which is not shown here. In contrast, for the other two cycles where $\Xi_{3,4} > \Xi_{1,2}$, positive work output is obtained, indicating their potential as quantum heat engines. Thus, we conclude that a GQOC with symmetric qubit-system coupling utilizing this particular working substance is not a promising candidate for a quantum heat engine. Moreover, as illustrated in Fig.\ref{fig2}, increasing the mean temperature $T_M$ of the reservoirs leads to an enhancement in the amount of positive work extracted in both the GQOC with asymmetric coupling ($W_A$) and QOC ($W_O$) cases. 
\begin{figure*}
\centering
\fbox{\includegraphics[trim={0 3.5cm 0 1cm},clip,width=\textwidth]{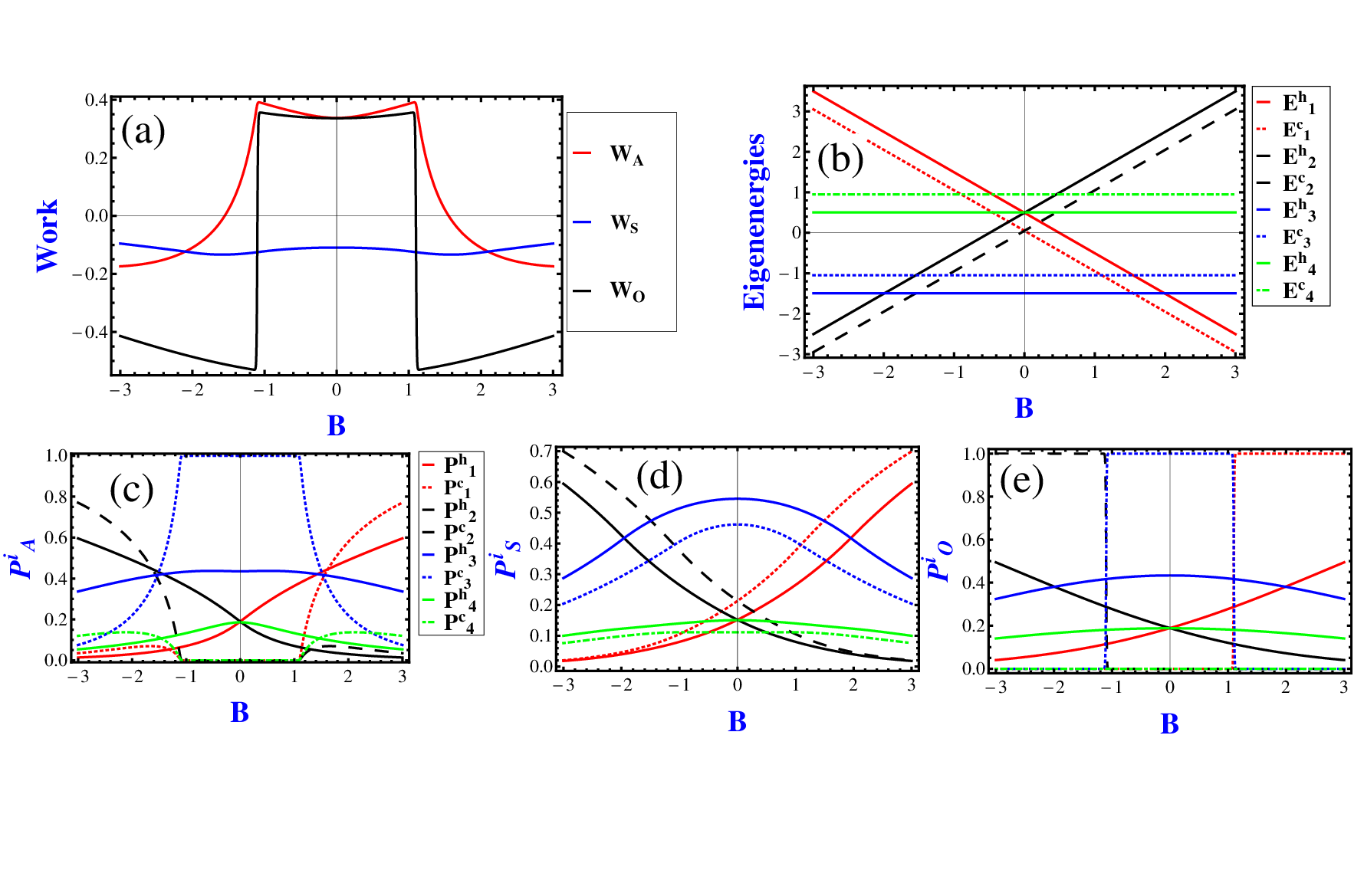}}
\caption{a) The work output $W$ (in units of $\hbar J$) as a function of the applied magnetic field $B$ for three different cycles. b) The eigenenergies of the system during the hot (h) and cold (c) isochoric stages of the quantum Otto cycle (equivalent to two non-equilibrium processes in the GQOC) as a function of the applied magnetic field $B$. The amount of system populations during the hot and cold isochoric stages as a function of $B$ are presented for c) GQOC with asymmetric coupling, d) GQOC with symmetric coupling, and e) quantum Otto cycle. The model parameters are set to $\Delta^{c}=0.10$, $\Delta^{h}=0.99$, $\kappa=0.05$, $T_M =1.2$ and $\Delta T=2T_M$, where all parameters are expressed in units of the interqubit coupling $J$.}
\label{fig3} 
\end{figure*}

To shed light on the question of why the condition $\Xi_{3,4} > \Xi_{1,2}$ cannot be satisfied in the GQOC with symmetric coupling while it can hold for the other two cycles, we need to delve deeper into the dynamics of each quantum state population in the hot and cold isochoric processes of the QOC and their equivalent processes in GQOC. Fig.\ref{fig3} illustrates the work, eigenenergies, and populations of all three cycles as a function of the applied magnetic field. In panel (a), the work output of the cycles is plotted against $B$. It is evident that only $W_A$ and $W_O$ exhibit positive work output, while the work output of $W_O$ sharply drops at $B_{cr}=\pm (\Delta^{c}+J)$, the work output of $W_A$ smoothly expandes over larger values of $|B|$. Panel (b) displays the eigenenergies $E_i^{h}$ and $E_i^{c}$ as a function of $B$, and it is apparent that a significant shift in energy dispersions occurs due to the change from $\Delta^c$ to $\Delta^h$. Panels (c)-(e) show the populations related to the four quantum states of the system in the two hot and cold isochoric processes (Processes 1-2 and 3-4 in GQOCs). Panel (b) shows that as the magnetic field $B$ changes, the ground state of the system transitions among $E_1$, $E_2$ and $E_3$. More importantly, when panel (c)-(e) are compared to panel (b), it is found that the probability of finding the system in its ground state is always the most likely outcome for all three cycles. This is due to the fact that at low temperatures (close to absolute zero), the population of the system follows the canonical ensemble distribution $\rho_{ii}(T)=\exp{(-\frac{E_i}{T})}$, and the system is predominantly in its ground state. As the temperature of the reservoirs is increased, the populations of the eigenstates $|\Phi_1\rangle$, $|\Phi_2\rangle$, $|\Phi_3\rangle$ and $|\Phi_4\rangle$ become more balanced due to state mixing. In our case, the temperatures of the reservoirs are relatively low ($T_M =1.2$), which explains the large values for the ground state population.

In the subsequent analysis, we focus on a specific region of Fig.\ref{fig3}, namely the interval where positive work output is observed, approximately at $-1.1<B<1.1$. By comparing panel (a) with (c) and(e), we note that in this region for both the GQOC with asymmetric coupling and Otto cases (panel (c) and (e)), the probability of finding the system in state $|\Phi_3\rangle$ at the end of the cold isochoric process (non-equilibrium \emph{processes (1-2)} in GQOC) is unity ($P_{3}^{c}=1$), while the probability of finding the system in any other state is zero ($P_{i}^{c}=0$). Moreover, we have $P_{3}^{h}<0.5$ and $P_{4}^{h}<0.2$, resulting in $\Xi_{3,4}>0$. Additionally, it is observed that $P_{1}^{h}>0$ and $P_{2}^{h}>0$, which yields $\Xi_{1,2}<0$ for the same cycles. Therefore, the condition of Eq.(\ref{W condition}) is automatically satisfied for these two cycles, indicating the occurrence of positive work output. In contrast, for the symmetric cycle (panel (d)), it is noted that $\Xi_{1,2}\geq0$ and $\Xi_{3,4}<0$ hold in the entire range of $B$, which is in complete contradiction to the condition of Eq. (\ref{W condition}). Therefore, the symmetric cycle does not exhibit positive work output. In order to gain further insights into this problem, it is crucial to examine the structure of the allowed transitions in hot and cold isochoric processes (non-equilibrium processes in GQOC).

\begin{figure*}
\centering
\fbox{\includegraphics[trim={0 0 0 0},clip,width=\textwidth]{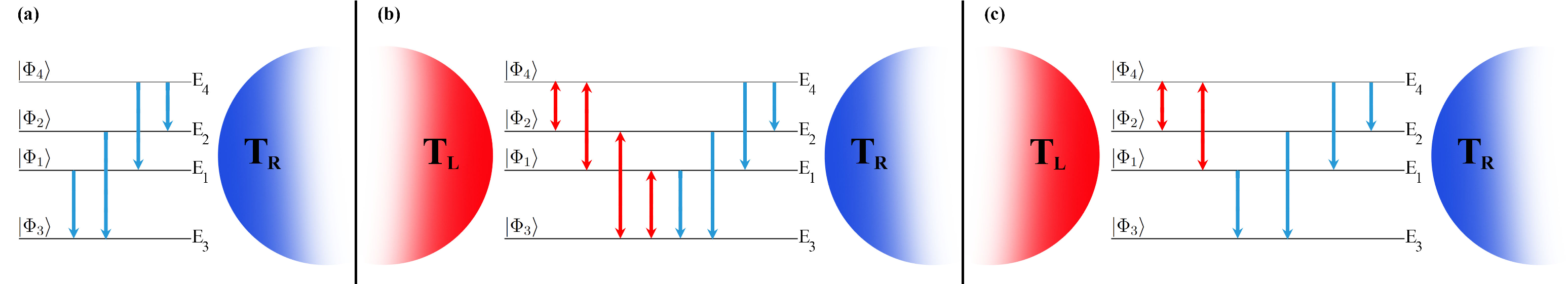}}
\caption{ The level structure of our four-level system at the end of process 1-2, showcasing the allowed transitions in the interaction with the hot and cold reservoirs. In (a) the traditional QOC is shown, while (b) and (c) depict the GQOC with symmetric and asymmetric coupling, respectively. The red and blue arrows indicate the permitted transitions in the interaction with the hot and cold reservoirs. }
\label{transitions} 
\end{figure*}

Fig.\ref{transitions} illustrates the energy levels of the system, mainly within the range of $-B_{cr}<B<B_{cr}$, and its allowed transitions during the interaction with hot and cold reservoirs at the end of the cold isochoric process (non-equilibrium \emph{processes (1-2)} in GQOC). Specifically, panels (a)-(c) represent the QOC, GQOC with symmetric coupling, and the GQOC with asymmetric coupling, respectively.

In this figure, the horizontal lines denote the energy levels of the system, and the arrows indicate the possible transitions that can occur between different energy levels during the interaction with the hot and cold reservoirs. The red and blue arrows represent transitions occurring during the interaction with the hot and cold reservoirs, respectively.

As previously demonstrated, the system's complete population at $|\Phi_3\rangle$ ($P_{3}^{c}=1$) within the range $-B_{cr}<B<B_{cr}$ plays a crucial role in fulfilling the condition of Eq.(\ref{W condition}).  In Appendix (\ref{AppA}), we discuss the main effect of the asymmetric coupling of the qubits with the reservoirs, which puts restrictions on the allowed transitions between the states induced by the environment. In the symmetric case ($\epsilon=0$), each reservoir can induce four different transitions, while in the asymmetric case ($\epsilon=1$), two sets of transitions $|\Phi_3\rangle\leftrightarrow|\Phi_1\rangle$ and $|\Phi_3\rangle\leftrightarrow|\Phi_2\rangle$ induced by the left reservoir are forbidden due to non-linear couplings.

To gain insight into the asymptotic behaviors of the eigenstate population in extreme conditions (i.e., at high temperature gradients), we analyze the emission (absorption) rates $\gamma_{ij}^{(\nu,e)}$ ($\gamma_{ij}^{(\nu,a)}$) given in Eq.(\ref{E10-a} - \ref{E10-h}) (see Appendix (\ref{AppA})). By fixing $T_L$ and taking the limit of $T_R\simeq0$ (the high temperature gradient limit), we obtain $\gamma_{13}^{(R,a)}=\gamma_{23}^{(R,a)}=0$, $\gamma_{13}^{(R,e)}=\frac{\kappa(B_{cr}-B)}{2}$, and $\gamma_{23}^{(R,e)}=\frac{\kappa(B_c+B)}{2}$. This indicates that at high temperature gradients and for $B\leq |B_{cr}|$, transitions from $|\Phi_3\rangle\rightarrow|\Phi_1\rangle$ and $|\Phi_3\rangle\rightarrow|\Phi_2\rangle$ are forbidden, while the reverse transitions induced by the interaction of the system with the cold reservoir occur. Furthermore, the transitions $|\Phi_3\rangle\leftrightarrow|\Phi_1\rangle$ and $|\Phi_3\rangle\leftrightarrow|\Phi_2\rangle$ induced by the interaction of the system with the hot reservoir are never allowed due to the inherent asymmetry. These physical reasons explain why the system with asymmetric coupling populates at the ground state $|\Phi_3\rangle$ in the high temperature gradient limit.

However, the explanation for $P_{3}^{c}=1$ in the case of the QOC is fundamentally different. Fig.\ref{transitions}, panel (a), illustrates the possible transitions induced by the cold reservoirs at the end of the cold isochoric process in the quantum Otto engine. It is evident from this panel that, for this specific situation, all allowed transitions occur in the form of emission. At the end of each isochoric stage, the system thermalizes with its surrounding thermal reservoir at temperature $T$, and its populations follow the Gibbs distribution $\rho_{ii}(T)=\exp{(-\frac{E_i}{T})}$. As previously mentioned, in the extreme conditions of high temperature gradient, $T_{cold}$ tends towards zero. Consequently, the entire system is populated at the ground state  $|\Phi_3\rangle$, and the approximately zero temperature environment is incapable of inducing any excitations in the energy levels of the system.

Finally, all permitted transitions of the system with symmetric couplings to the left and right thermal reservoirs are depicted in Fig.\ref{transitions}, panel (b). In contrast to the asymmetric case, all four possible transitions resulting from the interaction of the system with the environment are allowed for both hot and cold reservoirs. Therefore, in the high temperature gradient limit, where the temperature of the cold reservoir tends to zero, the hot reservoir with a finite temperature can now excite the system from the ground state to the upper states, resulting in a mixed state with $P_{3}^{c}\not=1$.\\

It is crucial to note that reversing the direction of the cycle does not allow for the extraction of positive work from the GQOC in the symmetric coupling scenario. According to Eqs.(\ref{def:1} - \ref{def:3}), reversing the cycle’s direction results in equations identical to those of the forward cycle, with only apparent differences in the occupation probabilities. However, since the steady-state solution of the Lindblad equation is unique, the steady-state occupation probabilities for both non-equilibrium stages remain identical in the forward and reverse cycles. Consequently, the net heat exchange and, thus, the work extraction remain unchanged in both directions, leading to zero net work output in the symmetric coupling scenario.

With this comprehensive analysis of the possible scenarios for extracting useful work, we can exclude the GQOC with symmetric coupling from further discussion, as it cannot function as a quantum heat engine. Therefore, we will concentrate on the two remaining cycles in the subsequent sections.

\subsection{\label{sec 4} Quantum Work and efficiency}

\begin{figure*}
\centering
\includegraphics[trim={0 2.5cm 0 6cm},clip,width=\textwidth]{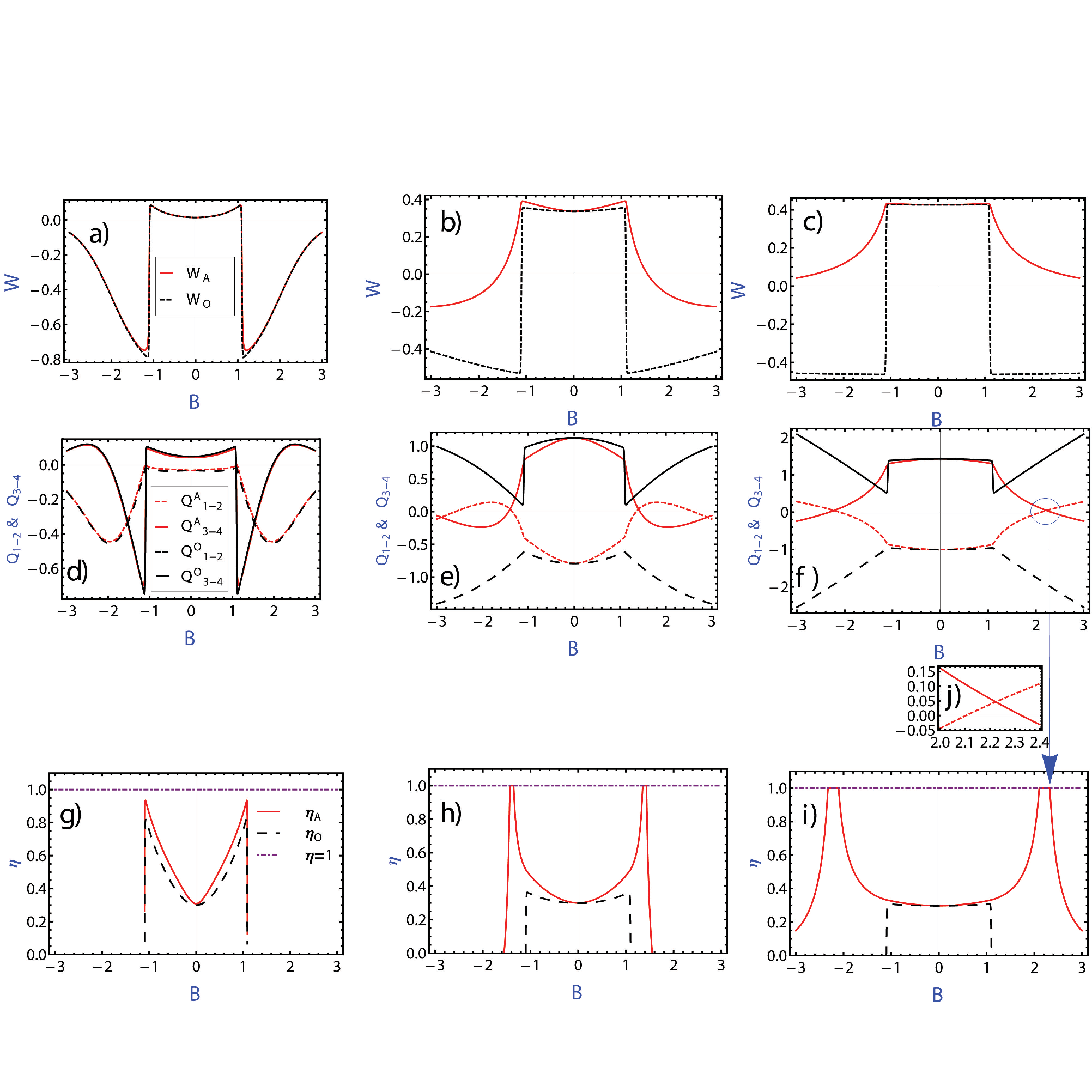}
\caption{Panels (a) to (c) the work output against the magnetic field $B$, panels (d) to (f) the amount of heat exchanged during the hot (processes (3-4) in GQOC) and cold (processes (1-2) in GQOC) isochoric processes and, panels (g) to (i) the efficiencies of the cycles versus $B$ for three different mean temperatures $T_M$. In the first column (panels (a), (d), and (g)), the mean temperature is $T_M=0.21$, while for the second column (panels (b), (e), and (h)), and the third column (panels (c), (f), (j), and (i)), $T_M=1.2$ and $T_M=6$, respectively. The model parameters are set to $\Delta^{c}=0.10$, $\Delta^{h}=0.99$, $\kappa=0.05$, and $\Delta T=2T_M$, where all parameters are expressed in units of the interqubit coupling $J$.}
\label{fig4} 
\end{figure*}

In this section, we explore the quantum work and efficiency for the two remaining cycles, i.e., the GQOC with asymmetric coupling and the QOC. Panels (a)-(c) of Fig.\ref{fig4} illustrate the work output against the magnetic field $B$, while panels (d)-(f) depict the amount of heat exchanged during the hot (processes (3-4) in GQOC) and cold (processes (1-2) in GQOC) isochoric processes. Finally, panels (g)-(i) present the efficiencies of the cycles versus $B$ for three different mean temperatures $T_M$. In the first column (panels (a), (d), and (g)), the mean temperature is $T_M=0.21$, while for the second column (panels (b), (e), and (h)), and the third column (panels (c), (f), (j), and (i)), $T_M=1.2$ and $T_M=6$, respectively.

As illustrated in Fig.\ref{fig4}, panels (a), (b) and (c), an increase in the base temperature of the reservoirs results in a higher amount of extracted work for both the GQOC with asymmetric coupling and the QOC. Additionally, it expands the domain of positive work exclusively for the GQOC. It is important to note that, given we are operating in a high temperature gradient regime ($\Delta T = 2T_M$), the mean temperature depends solely on the hot bath temperature $T_{h}$, as in this regime $T_{c} \rightarrow 0$. At low base temperatures, the performance of the QOC and the GQOC with asymmetric coupling are almost the same in terms of work output (as seen in panel (a) of Fig.\ref{fig4}) because their values of heat exchanged are approximately the same (as seen in panel (d)). However, the efficiency of the GQOC with asymmetric coupling is slightly higher than the QOC (as seen in panel (g)). Additionally, when the base temperature of the reservoirs tends to zero and $T_{M} \approx 0$, then both reservoirs could be considered as one reservoir with low temperature. Thus, the GQOC turns into the QOC.

As shown in Fig.\ref{fig4} panels (b) and (c), for higher base temperatures ($T_M=1.2$ and $T_M=6$, respectively), there is a clear difference between the work outputs of the two cycles. As the mean temperature $T_M$ increases, the extracted work from the GQOC with asymmetric coupling broadens over magnetic field $B$, while the amount of maximum work output approaches the same saturation value as the QOC.

To understand why both cycles exhibit the same maximum values within a specific domain of $B$, we focus on the relevant region where $B\leq |B_{cr}|$. We first examine the performance of the QOC in this region. According to Eq.(\ref{w:def}-c), the work output depends on $P_{ij}^{h/c}$ and $E_i^{h/c}$, Since changes in the $T_{M}$ do not affect the energy dispersion of the system, the only variables that play a crucial role in the amount of heat absorbed or released, and consequently the work output, in panels (b) and (c) of Fig.\ref{fig4} are the probabilities $P_{ij}^{h/c}$. As previously mentioned, when the system is thermally equilibrated with a reservoir at temperature $T$, its probability ratio follows the Gibbs distribution:

\begin{eqnarray}
\label{P ratio}
\frac{P_i}{P_j}=\exp{(\frac{E_j-E_i}{T})}.
\end{eqnarray}

Accordingly, at the end of the cold isochoric stage, when the system thermalizes with the cold reservoir, the state with lower energy will always have a higher probability of being occupied than the state with higher energy. Hence, as ($T_{c}\rightarrow0$), the probability distribution of the system becomes $P_{3}^{c}=1$ and $P_i^{c}=0$ for $i=1,2,4$, with the system being almost exclusively populated on the ground state $|\Phi_3\rangle$. On the other hand, at the end of the hot isochoric stage when the system is in thermal equilibrium with the hot reservoir, the ratio $\frac{P_i}{P_j}$ tends towards 1 as the base temperature of the hot reservoir increases to higher levels, and the density matrix of the system becomes a maximally mixed state represented by $\rho=\frac{1}{d}\mathbb{I}$, where $d$ is the dimension of the Hilbert space and $\mathbb{I}$ is the identity matrix. Consequently, further increases in temperature result in negligible changes in the occupation probabilities, and hence, the values of $Q_{1-2}$, $Q_{3-4}$, and $W_O$ remain fixed.

Now, we investigate the asymmetric GQOC within the specific region where $B\leq |B_{cr}|$. According to the discussion in Sec. \ref{sec 3}, at the end of non-equilibrium processes (1-2), the steady-state probability distribution of the system is determined to be $P_{3}^{c}=1$,which is identical to the distribution observed at the end of the cold isochoric process in the QOC. While, the populations of the states at the end of the non-equilibrium processes (3-4) tend to establish a fully mixed state as the mean temperature reaches higher values, (see Appendix (\ref{AppB})). By considering the population values ($P_3^c=1$ and $P_1^h$, $P_2^h$, $P_3^h$, and $P_4^h$ being approximately equal) and applying Eq.(\ref{w:def}-c), the maximum work output for both cycles can be calculated as follows:

\begin{align}
W_{\text{max}} &= \frac{\Delta^h-\Delta^c}{2}\left[P_3^c-P_3^h+P_1^h+P_2^h-P_4^h\right] \notag \\
W_{\text{max}} &\cong \frac{\Delta^h-\Delta^c}{2}.
\label{eq:wmax}
\end{align}
 
The heat transfer analysis depicted in panels (d)-(f) illustrates that for the region where QOC generates positive $W$, the amount of heat input to the system is greater than the amount of heat output to the cold reservoir, a condition that is reflected in $Q_{3-4}>|Q_{1-2}|>0$. This condition, however, does not necessarily apply to the GQOC. As in GQOC the system is not thermally equilibrated with its environment during both non-equilibrium processes, allowing for the exchange of arbitrary amounts of heat.

In panels (g) to (i), we employ the aforementioned condition to plot $\eta_O$, while only Eq.(\ref{w:def}-d) is used to obtain $\eta_A$. These panels demonstrate that the efficiency of the asymmetric GQOC increases with the mean temperature and surprisingly reaches its maximum value of 1 (or 100\%), while the efficiency of the Quantum Otto cycle decreases to a saturation value. In high mean temperatures, where the graphs show a clear separation between the curves, the values of $Q_{1-2}$ and $Q_{3-4}$ for QOC become exclusively negative and positive, respectively, whereas in GQOC, both $Q$s can take on either positive or negative values. Panel (j) in Fig.\ref{fig4} illustrates that the maximum efficiency for GQOC occurs in regions where both $Q$s are positive. This indicates that, for appropriate values of parameters, it is possible to absorb heat from the environment during both the 1-2 and 3-4 non-equilibrium stages and convert all absorbed heat into useful work, in accordance with the conservation law of energy in a cycle. Achieving the Carnot limit of efficiency or even 100\% efficiency through the conventional QOC with thermal reservoirs has been reported in previous studies \cite{Must1, Z4, Allah1, Must2}, with discussions on the possibility of attaining such efficiencies. In all these studies, it was shown that such efficiencies are typically achieved when the work output approaches zero. 
On the other hand, in some other studies, maximum Carnot efficiency (or even efficiencies exceeding Carnot) was achieved in QOC with non-zero work output ($W\ne 0$), by employing non-equilibrium reservoirs, such as squeezed thermal baths \cite{Aba, Rob, Almeida}.\\
 Interestingly, our GQOC demonstrates that, by using non-equilibrium reservoirs and asymmetric coupling, 100\% efficiency can be achieved even when the work output remains significantly greater than zero ($W>0$). In our GQOC this arises from the fact that both $Q_{1-2}$ and $Q_{3-4}$ in $W=Q_{1-2}+Q_{3-4}$ are positive and thus, according to Eq.(\ref{w:def}-d), the efficiency becomes $\eta=1$. In the following section, we will determine the permissible ranges of system and environment parameters where $Q_{1-2}\ge 0$ and $Q_{3-4}\ge 0$ hold simultaneously.

\subsection{\label{subsec eta cond}Conditions for 100\% Efficiency}
The necessary conditions to achieve maximum efficiency of 1 can be determined by solving the following inequality
\begin{eqnarray}
 \Xi_{3,4}-\Xi_{1,2}>\frac{Q_{1-2}}{W_{max}}>0.
\label{eta-condition}
\end{eqnarray}

\begin{figure}
  \centering
  \fbox{\includegraphics[width=0.8\columnwidth, trim={0.5cm 1cm 0.5cm 2.2cm}, clip=true]{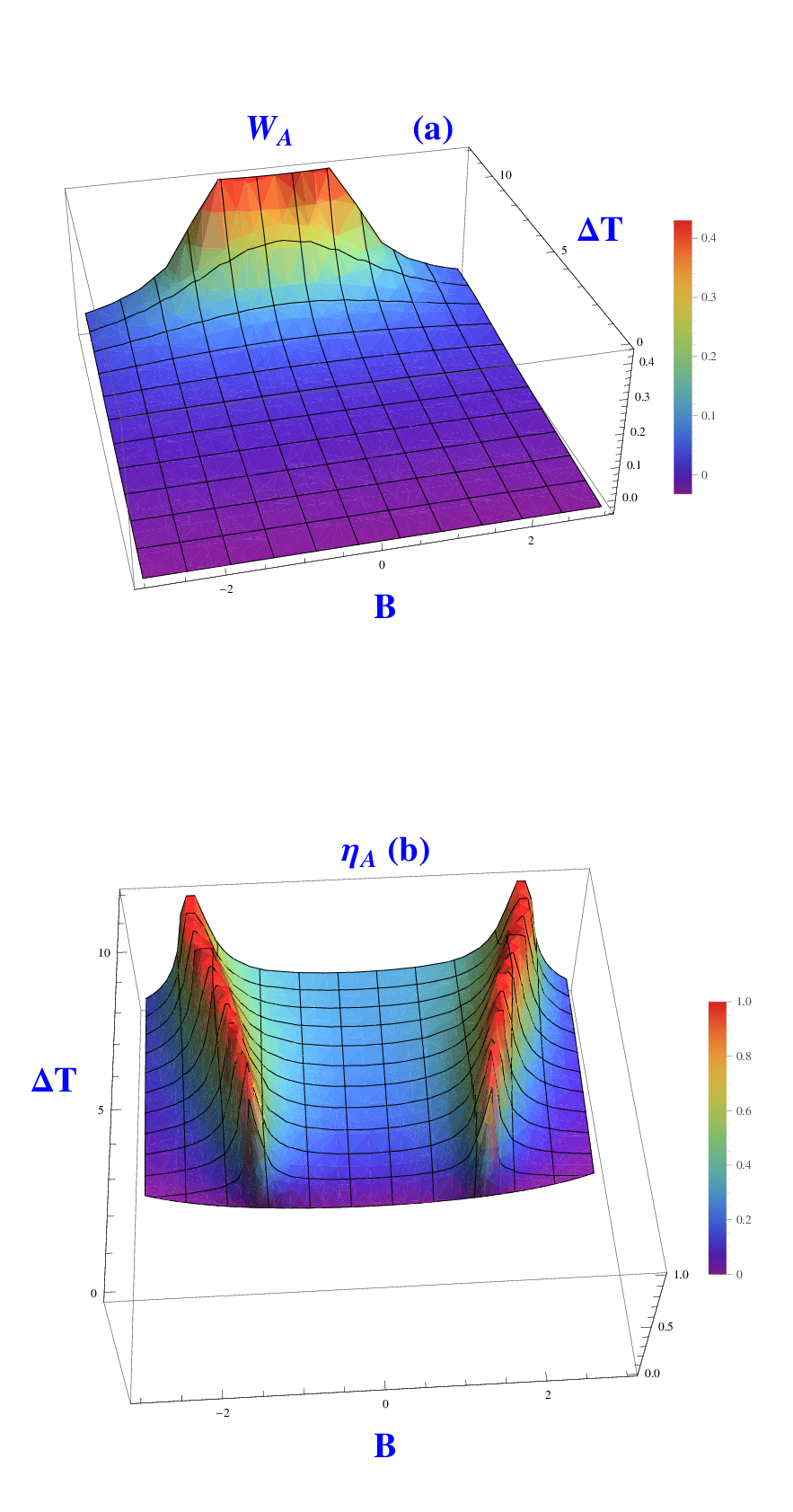}}
  \caption{The work output (a) and the efficiency (b) of the asymmetric GQOC as functions of $B$ and $\Delta T$. The model parameters are set to $\Delta^{c}=0.10$, $\Delta^{h}=0.99$, $\kappa=0.05$, and $T_M=6$, where all parameters are expressed in units of the interqubit coupling $J$. }
  \label{fig5}
\end{figure}

This inequality not only satisfies the positive work condition described in Eq.\ref{W condition}, but also imposes an additional, more stringent condition. Specifically, to achieve 100\% efficiency, the sum of the population differences of the entangled states during the cycle must exceed those of the unentangled states by a positive term $\frac{Q_{1-2}}{W_{max}}$. In order to explore for which choice of parameters this equation is true, we have plotted the work output and the efficiency of the asymmetric GQOC as functions of $B$ and temperature gradient of the reservoirs in Fig.\ref{fig5}.

In panel (b) of this figure, the regions where the GQOC with asymmetric coupling achieves an efficiency of 1 are displayed. As shown, there are two symmetric tracks around $B=0$ where the conditions of Eq.\ref{eta-condition} are met, resulting in an efficiency of 1. Additionally, even outside the maximum temperature gradient regime, i.e., $\Delta T < 2T_M$, the maximum efficiency of 1 can still be achieved. However, as shown in panel (a), in regimes close to $\Delta T \rightarrow 2T_M$, the work output is higher compared to other areas. 
 Therefore, it is preferable to focus on the regime where $\Delta T = 2T_M$ as it allows for both maximum work output and maximum 100\% efficiency, although they do not occur at the same value of $B$.\\

\subsection{\label{second law}Consistency with the second law of thermodynamics}
An important question that naturally arises is whether achieving 100\% efficiency, as presented in our results, is consistent with the second law of thermodynamics. For conventional engines operating between two reservoirs (hot and cold), the second law requires that heat is absorbed from the hot reservoir, and it is impossible to convert all the absorbed heat into work without releasing some to the cold sink. However, in the GQOC, the working substance interacts simultaneously with both the hot and cold reservoirs at each Non-equilibrium stage. As demonstrated in Section \Romannum{3}-A, the corresponding quantum heats in both non-equilibrium stages are positive. Therefore, the conventional interpretation of the second law for standard QOCs does not directly apply to the GQOC. Another formulation of the second law, based on entropy, states that the entropy production rate must remain positive in any quantum process \cite{Breuer}. In this section, we will demonstrate that the GQOC fully complies with this statement of the second law of thermodynamics. According to the second law of thermodynamics, the entropy of an open system, whether classical or quantum, changes according to a balance equation, which can be expressed in the form: \cite{Breuer,Paternostro2019, Abah2024}:\\

\begin{eqnarray}
\frac{dS}{dt}=\Pi - \Phi,
\label{SCL}
\end{eqnarray}

\begin{figure*}
  \centering
  \includegraphics[width=\textwidth, trim={0cm 0cm 0cm 0cm}, clip=true]{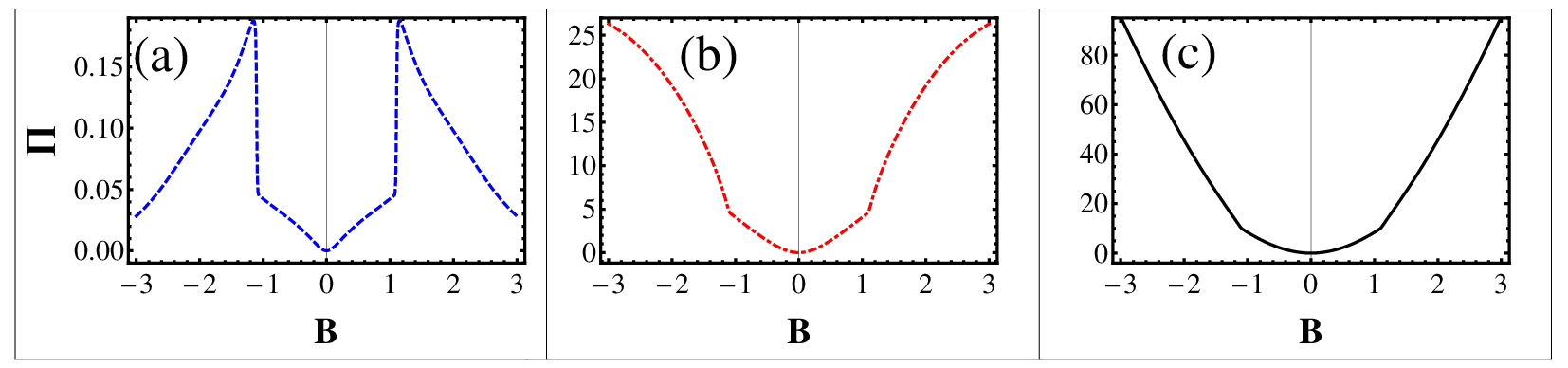}
  \caption{Entropy production rate of the GQOC as a function of magnetic field $B$, panels (a) to (c) cprespond to $T_{M}=0.21$,  $T_{M}=1.2$ and  $T_{M}=6$ respectively.  All other parameters are consistent with those used in Fig.\ref{fig4}.}
  \label{figEP}
\end{figure*}

where $\Pi$ denotes the irreversible entropy production rate, and $\Phi$ represents the entropy flux, indicating the flow of entropy per unit time between the open system and the reservoirs. If $\Phi>0$, entropy flows from the system to the reservoirs, and if $\Phi<0$, it flows from the reservoirs to the system. When the system reaches a nonequilibrium steady state characterized by $\frac{dS}{dt}=0$, it follows that $\Pi=\Phi$. In this case, the system continuously generates entropy, all of which flows into the reservoirs. The second law of thermodynamics mandates $\Pi \ge 0$, where $\Pi= 0$, holds only when the system is at equilibrium \cite{Breuer}.\\
Now, we aim to demonstrate that ($\Pi > 0$) always holds for our GQOC.
Entropy flux is related to the heat current as follows \cite{Sen2022}:

\begin{eqnarray}
\Phi=-\frac{\boldmathcal{Q}_{L}}{T_{L}}-\frac{\boldmathcal{Q}_{R}}{T_{R}},
\label{Phi}
\end{eqnarray}

where $\boldmathcal{Q_{\nu}}=\frac{dQ_{\nu}}{dt}$, $(\nu=L,R)$ is the quantum heat current and $T_{\nu}$ is the temperature of the reservoir $\mathcal{R}_{\nu}$. The heat current can be unambiguously defined as the time variation of the average energy going through the system, given by the following expression:

\begin{equation}
\begin{aligned}
\frac{d\langle H\rangle}{dt} &=Tr\{\mathcal{D}_{L}(\rho)H_{S}\}+Tr\{\mathcal{D}_{R}(\rho)H_{S}\} \\&=(\boldmathcal{Q}_{L}(t)+\boldmathcal{Q}_{R}(t)),
\end{aligned}
\label{J1}
\end{equation}

where $\boldmathcal{Q}_{\nu}(t)=Tr\{\mathcal{D}_{\nu}(\rho)H_{S}\}$ represent the heat current from the reservoir $\mathcal{R}_{\nu}$ into the system, and $\mathcal{D}_{\nu}(\rho)$ is the dissipator associated with bath $\mathcal{R}_{\nu}$, see Appendix (\ref{AppA}). Substituting $\mathcal{D}_{\nu}(\rho)$ from Eq.(\ref{Dis}) into Eq.(\ref{J1}) we derive the following expression for the quantum heat current:\\

 \begin{equation}
\begin{aligned}
\boldmathcal{Q}_{\nu}(t)&=(E_{3} - E_{1})\Gamma_{13}^{\nu}(t)+(E_{4} - E_{1})\Gamma^{\nu}_{14}(t)\\&+(E_{3} - E_{2})\Gamma^{\nu}_{23}(t)+(E_{4} - E_{2})\Gamma^{\nu}_{24}(t),
\end{aligned}
\label{J2}
\end{equation}

where $\Gamma_{ij}^{\nu}(t)=\gamma_{ij}^{(\nu,e)}\rho_{ii}(t)-\gamma_{ij}^{(\nu,a)}\rho_{jj}(t)$ represents the net decay rate from the state $|\Phi_{i}\rangle$ to $|\Phi_{j}\rangle$. Here, $\gamma_{ij}^{(\nu,e)}$ and $\gamma_{ij}^{(\nu,a)}$ correspond to the emission and absorption rates between the states $|\Phi_{i}\rangle$ to $|\Phi_{j}\rangle$, respectively. It is important to note that, in general, $\boldmathcal{Q}_{L}(t)\ne \boldmathcal{Q}_{R}(t)$. However, at steady state, the total heat current flowing through the system is zero, i.e., $\boldmathcal{Q}_{L}(t=\infty)+ \boldmathcal{Q}_{R}(t=\infty)=0$. During the two adiabatic processes, since the population distributions remain constant, the entropy production rate is zero. Therefore, to calculate $\Pi$ for the entire cycle, it suffices to evaluate $\Pi$ during the two non-equilibrium stages.

In Fig.\ref{figEP}, the entropy production rate $\Pi$ of the GQOC is shown as a function of the magnetic field $B$ for different values of the mean temperature $T_{M}$. Notably, $\Pi$ remains positive across the entire parameter range, providing strong evidence for the consistency of our proposed quantum engine with the second law of thermodynamics.\\

\section{\label{sec possible im} A Possible Implementation}
\begin{figure}
  \centering
  \fbox{\includegraphics[width=1\columnwidth, trim={0cm 0cm 0cm 0cm}, clip=true]{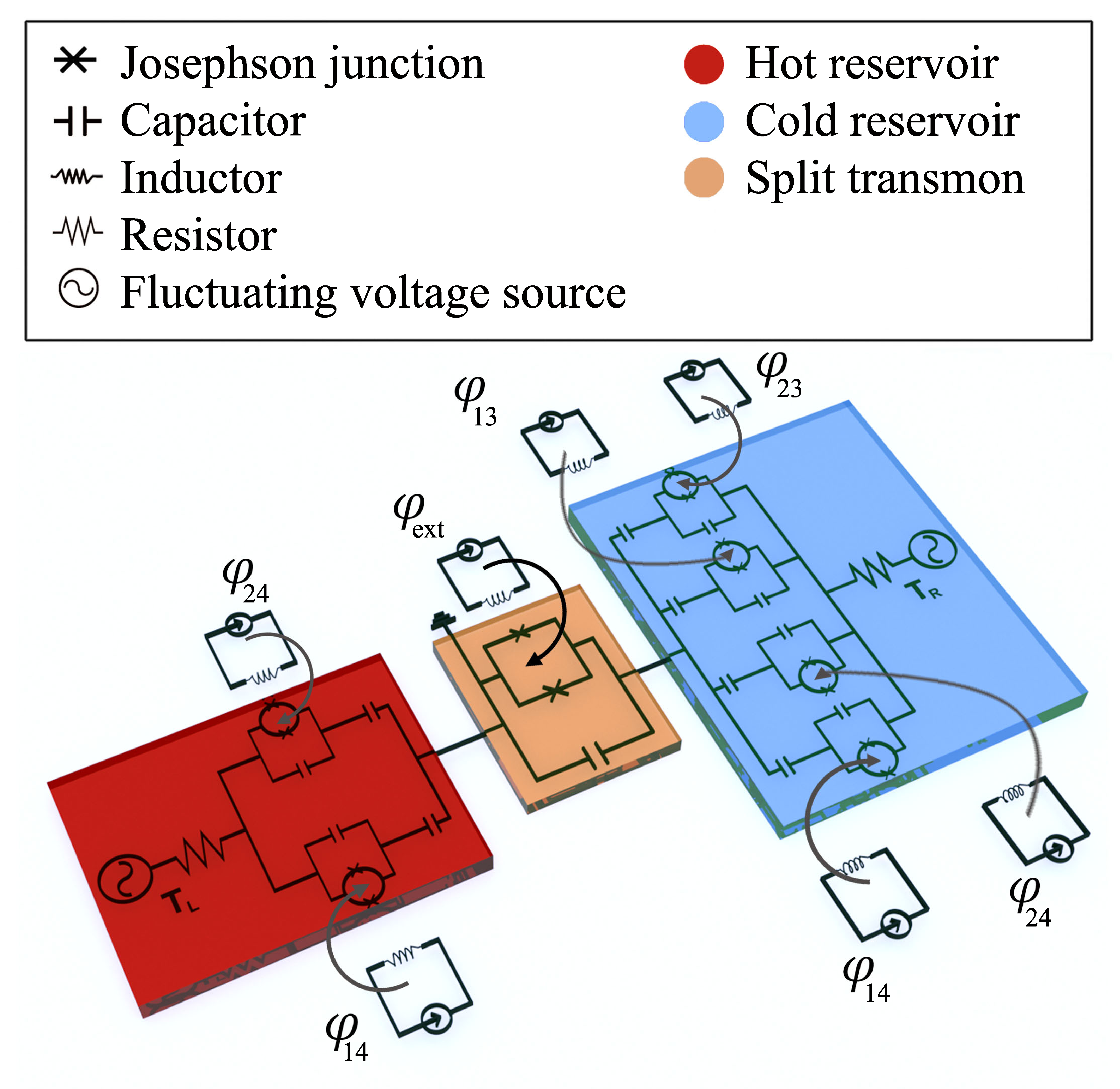}}
  \caption{Schematic of the implementation of GQOC with asymmetric coupling in circuit QED. The thermal reservoirs are represented by sets of LC circuits functioning as bandpass filters, coupled to noisy resistors. The left reservoir is composed of two LC circuits, while the right reservoir consists of four LC circuits. A tunable split transmon qudit is naturally coupled to the thermal reservoirs through the electromagnetic noise generated by the thermal agitation of Cooper pairs. By adjusting the external flux $\upvarphi_{ij}$ through the superconducting quantum interference loop of the qudit, the energy levels of the split transmon can be shifted. Additionally, tuning the external flux through the loops of the LC circuits allows for the adjustment of the resonant frequencies, enabling the coupling or decoupling of the system from the reservoirs. }
  \label{fig6}
\end{figure}

In this section, we will discuss potential methods for physically realizing the introduced QOC and GQOC. To achieve this, careful engineering of the isochoric (non-equilibrium stages in GQOC) and adiabatic processes is necessary. To model the four level system with desired transitions depicted in Fig.\ref{transitions} panel (c) at the non-equilibrium stage of the GQOC, we use a split transmon qudit \cite{Kran} that is coupled to two thermal bath circuits, as shown in Fig.\ref{fig6}. A split transmon qudit is a system that consists of two Josephson junctions in a superconducting quantum interference loop, allowing the transition frequency to be tuned by applying an external magnetic field that threads a flux through the superconducting quantum interference loop. The thermal bath circuit includes a voltage noise generator coupled to a series of band-pass filters.

A voltage noise generator is typically modeled by an effective circuit consisting of a real resistor with a fluctuating voltage source. Similarly, a band-pass filter can be represented by a LC circuit that can be tuned via external flux to allow only the voltage noise near its resonance frequency to pass through, while blocking all others. The quantum theory of resistors \cite{Clerk} can be used to describe the Hamiltonian of the generated voltage noise in the thermal circuit, which can be expressed as $H = \sum_{k} \omega_j b_{k}^{\dagger} b_k$. The temperature of the thermal bath circuit can be controlled by modulating the noise level of the resistor, and the circuit generates a continuous spectrum of voltage noise due to the thermal agitation of Cooper pairs. In the model of the isochoric stage of the QOC, band-pass filters are not required, as the system will naturally thermalize with the thermal bath. Now, by tuning the LC frequency around the $\Omega=E_i - E_j$, the qudit interacts with bath by exchanging energy in the form of photons of energy $\hbar \Omega$ associated with one of the level spacing of the qudit. This interaction can be described by the following Hamiltonian
\begin{eqnarray}
 H_{int}=\hbar \alpha_k \left(\sigma_{-}b_{k}^{\dagger}+\sigma_{+}b_k \right),
\label{Hint}
\end{eqnarray}
where $\alpha_k$ denotes the coupling constant between the qudit and the $k$th mode of the bath, and $\sigma_{+}$ ($\sigma_{-}$) is the raising (lowering) operator associated with the eigenstates $|\Phi_i\rangle$ and $|\Phi_j\rangle$. The qudit and bath circuits are connected through a capacitor which tunes the qudit-bath coupling strength, which is important to obtain a deeper understanding of the dynamics of the qudit. Based on the theory of open quantum systems, the qudit-reservoir coupling also plays a control role in the dynamics of the system as well as relaxation into the steady state. The only adjustable control parameter during an isochoric or non-equilibrium stage is the thermalization time or steady-state relaxation time which can be controlled either by adjusting the capacitive coupling strength or by tuning the filter frequencies. Here we assume that the capacitive coupling strength is so weak that it does not perturb the qudit levels, ensuring the avoidance of correlations between the qudit and the bath. In the weak-coupling regime, the dynamics along the isochoric or non-equilibrium stages is approximately described by the following Born-Markov master equation:
\begin{widetext}
\begin{align}
\dot{\rho}=&-i\left[H_s,\rho\left(t\right) \right] + \sum_{\Omega_{ij}>0} \mathcal{J}\left(\Omega_{ij}\right)\left(1+n\left(\Omega_{ij}\right)\right) \left[\mathcal{V}\left(\Omega_{ij}\right) \rho\left(t\right) \mathcal{V}^{\dagger}\left(\Omega_{ij}\right) -  \frac{1}{2} \left\{\mathcal{V}^{\dagger}\left(\Omega_{ij}\right)\mathcal{V}\left(\Omega_{ij}\right),\rho\left(t\right) \right\} \right] \notag \\
&+ \mathcal{J}\left(\Omega_{ij}\right)\left(n\left(\Omega_{ij}\right)\right)\left[\mathcal{V}^{\dagger}\left(\Omega_{ij}\right)\rho\left(t\right)\mathcal{V}\left(\Omega_{ij}\right)- \frac{1}{2}  \left\{\mathcal{V}\left(\Omega_{ij}\right)\mathcal{V}^{\dagger}\left(\Omega_{ij}\right),\rho\left(t\right) \right\} \right],
\label{ex-ro-dot}
\end{align}
\end{widetext}

where $H_s$ and $\rho(t)$ denote the Hamiltonian and reduced density matrix of the qudit, respectively. The term $n(\Omega_{ij}) = \left[e^{\frac{\hbar\Omega_{ij}}{k_{B}T}}-1 \right]^{-1}$ represents the thermal mean value of the number of excitations in the reservoir at the frequency $\Omega_{ij}$. The voltage noise spectrum of a thermal bath circuit is denoted by $\mathcal{J}(\Omega_{ij})$, as given by:

\begin{align}
\mathcal{J}(\Omega_{ij})=&\left[1+Q_{LC}^{2}\left(\frac{\Omega_{ij}}{\Omega_{LC}}-\frac{\Omega_{LC}}{\Omega_{ij}} \right)^{2} \right]^{-1}\frac{2\hbar\Omega_{ij}}{R\left(1-e^{\frac{\hbar\Omega_{ij}}{k_{B}T}} \right)},
\end{align}

where, $T$ represents the temperature of the bath, $R$ is the resistance of the circuit resistor, and $\Omega_{LC} = 1/\sqrt{LC}$ and $Q_{LC}=\sqrt{LC}/R$ are the resonance frequency and the quality factor of the $LC$-circuit, respectively.
During the adiabatic process, the filter circuit is tuned far from the frequency $\Omega_{ij}$ to isolate the qudit from the baths. Additionally, a slowly varying magnetic flux is applied to the qudit, resulting in changes in its level spacing. This process is described by a unitary evolution with the control time $\tau_{adi} \gg 1/E$, where $E$ is a typical energy scale for the qudit.

\section{\label{sec 5} Summary}
In conclusion, this study examined three distinct designs of quantum thermal engines that utilized a two-qubit Heisenberg XXZ system as the working substance. A novel GQOC was introduced, offering two different configurations for coupling the two-qubit system to thermal reservoirs. The performance of these configurations was compared with that of the conventional QOC through a comprehensive analysis. By establishing the essential condition for achieving positive work extraction, it was demonstrated that only the GQOC with an asymmetric configuration and the QOC were capable of generating positive work. The positive work condition was further investigated for all three cycles, taking into account the allowed transitions during the isochoric and non-equilibrium stages specific to the QOC and GQOC, respectively. Additionally, the work output, heat exchange, and cycle efficiencies were thoroughly examined, leading to the determination of the maximum achievable work output. Significantly, it was discovered that the GQOC has the potential to achieve 100\% efficiency, representing a notable advantage over the QOC, which does not exhibit this behavior. It is worth emphasizing that the GQOC represents a novel class of quantum engines, where achieving 100\% efficiency does not require the work output to approach zero, as is the case in some conventional QOCs. This advantage can be attributed to the novel quantum cycle, the asymmetric coupling configuration of the system to the reservoirs, and the judicious selection of relevant parameters. Furthermore, the crucial condition for attaining optimal performance was derived analytically. Finally, a proposed experimental setup utilizing superconducting quantum circuits was presented as a means to implement both the GQOC and QOC in practical applications.

\begin{acknowledgments}
This work was not supported by any funding.
\end{acknowledgments}

\appendix

\section{\label{AppA} Solution of the Quantum Master Equation and Determination of Transition Rates}

Within the framework of the Born-Markov approximation and under the assumption that the reservoirs are in a Gibbs state at a finite temperature $T_{i}$, the dynamics of the system can be effectively characterized by employing the following master equation\\

\begin{eqnarray}
\label{E6}
\dot{\rho}(t)=-i[H_s,\rho(t)]+\mathcal{L}_{L}^{(1)}(\rho)+\\ \nonumber
\sum_{\nu=L,R}\mathcal{L}_{\nu}^{(2)}(\rho)+\mathcal{L}^{(1,2)}_{L}(\rho),
\end{eqnarray}
where $\rho(t)$ represents density matrix of the system. The Lindblad operator $\mathcal{L}{\nu}^{(k)}(\rho)$ corresponds to the dissipation arising from the coupling of qubit $k$ to reservoir $R_{\nu}$. Additionally, $\mathcal{L}_{\nu}^{(1,2)}(\rho)$ accounts for the dissipation resulting from the common coupling of qubits 1 and 2 to the left reservoir. The explicit expressions for the Lindblad operators are given by:\\

\begin{widetext}
\begin{align}
\label{E7}
& \mathcal{L}_{\nu}^{(k)}(\rho)= \sum_{\omega>0}\mathcal{J}_{\nu}^{(k)}(\omega)
(1+n_{\nu}(\omega))\left[V_{\nu}^{(k)}(\omega)\,\rho(t)V^{(k)\dag}_{\nu}(\omega)
-\frac{1}{2}\{V^{(k)\dag}_{\nu}(\omega)V_{\nu}^{(k)}(\omega),\rho(t)\}\right] \nonumber \\
&\quad + \sum_{\omega>0}\mathcal{J}_{\nu}^{(k)}(\omega)n_{\nu}(\omega)\left[V^{(k)\dag}_{\nu}(\omega)\,\rho(t)V_{\nu}^{(k)}(\omega)
-\frac{1}{2}\{V_{\nu}^{(k)}(\omega)V^{(k)\dag}_{\nu}(\omega),\rho(t)\}\right], \nonumber  \\
&\mathcal{L}_{L}^{(1,2)}(\rho)= \sum_{\omega>0}\mathcal{J}_{L}^{(1,2)}(\omega)
(1+n_{L}(\omega))\left[V_{L}^{(1)}(\omega)\,\rho(t)V^{(2)\dag}_{L}(\omega)
-\frac{1}{2}\{V^{(2)\dag}_{L}(\omega)V_{L}^{(1)}(\omega),\rho(t)\}\right] \nonumber \\
&\quad + \sum_{\omega>0}\mathcal{J}_{L}^{(1,2)}(\omega)n_{L}(\omega)\left[V^{(1)\dag}_{L}(\omega)\,\rho(t)V_{L}^{(2)}(\omega)
-\frac{1}{2}\{V_{L}^{(2)}(\omega)V^{(1)\dag}_{L}(\omega),\rho(t)\}\right] \nonumber  \\
 &\quad + \sum_{\omega>0}\mathcal{J}_{L}^{(1,2)}(\omega)
(1+n_{L}(\omega))\left[V_{L}^{(2)}(\omega)\,\rho(t)V^{(1)\dag}_{L}(\omega)
-\frac{1}{2}\{V^{(1)\dag}_{L}(\omega)V_{L}^{(2)}(\omega),\rho(t)\}\right] \nonumber  \\
 &\quad + \sum_{\omega>0}\mathcal{J}_{L}^{(1,2)}(\omega)n_{L}(\omega)\left[V^{(2)\dag}_{L}(\omega)\,\rho(t)V_{L}^{(1)}(\omega)
-\frac{1}{2}\{V_{L}^{(1)}(\omega)V^{(2)\dag}_{L}(\omega),\rho(t)\}\right],  \\
& V_{\nu}^{(k)}(\omega) = \sum_{\omega>0}|\Phi_j\rangle\langle \Phi_j|\sigma_x^{(k)}|\Phi_i\rangle\langle \Phi_i|, 
\label{E7b}
\end{align}
\end{widetext}

For all eigenfrequencies $\omega=\omega_{ij}=E_i- E_j>0$ corresponding to transitions $|\Phi_i\rangle\leftrightarrow|\Phi_j\rangle$. Here, $n_{\nu}(\omega)=\left[e^{\frac{\omega}{T_{\nu}}}-1\right]^{-1}$ represents the thermal mean value of the number of excitations in reservoir $R_{\nu}$ at frequency $\omega$, and $\mathcal{J}_{\nu}^{(k)}(\omega)$ is the spectral density of that reservoir. The transition operators $V_{\nu}^{(k)}(\omega_{ij})$ and $V_{\nu}^{\dagger(k)}(\omega_{ij})$ correspond to processes where the system loses or gains energy from reservoir $R_{\nu}$ through the relaxation or excitation of qubit $k$, respectively. These operators are eigenoperators of $H_S$ and satisfy $V_{\nu}^{(k)}(\omega_{ij})=V_{\nu}^{\dagger(k)}(-\omega_{ij})$ and $[H_S,V_{\nu}^{(k)}(\omega_{ij})]=-\omega_{ij}V_{\nu}^{(k)}(\omega_{ij})$. In the following, we assume that the spectral coupling densities follow an ohmic form, $\mathcal{J}_{\nu}^{(k)}(\omega)=\kappa\omega$, where $\kappa$ is a dimensionless constant. Consequently, the collective damping rate becomes $\mathcal{J}_{\nu}^{(1,2)}(\omega)=\sqrt{\mathcal{J}_{\nu}^{(1)}(\omega)\mathcal{J}_{\nu}^{(2)}(\omega)}=\kappa\omega$. We note that the validity of the Markovian approximation is limited to the small $\kappa$ regime. By using the laddering relation $\sigma_x^{(k)}|\pm\rangle=|\mp\rangle$, Eq.(\ref{HRS}) determines the explicit forms of the non-vanishing transition operators $V_{\nu}^{(k)}(\omega_{ij})$.\\
\begin{widetext}
\begin{align}
\label{E8}
&\hspace{-1.7cm}V_{L}^{(1)}(\omega_{13})=-\frac{1}{\sqrt{2}}|\Phi_3\rangle\langle\Phi_1|,\quad\,\,V_{L}^{(2)}(\omega_{13})=V_{R}^{(2)}(\omega_{13})=\frac{1}{\sqrt{2}}|\Phi_3\rangle\langle\Phi_1|\\
&\hspace{-1.7cm}V_{L}^{(1)}(\omega_{14})=\frac{1}{\sqrt{2}}|\Phi_4\rangle\langle\Phi_1|,\quad\,\,V_{L}^{(2)}(\omega_{14})=V_{R}^{(2)}(\omega_{14})=\frac{1}{\sqrt{2}}|\Phi_4\rangle\langle\Phi_1|\\
&\hspace{-1.7cm}V_{L}^{(1)}(\omega_{23})=-\frac{1}{\sqrt{2}}|\Phi_3\rangle\langle\Phi_2|,\quad\,\,V_{L}^{(2)}(\omega_{23})=V_{R}^{(2)}(\omega_{23})=-\frac{1}{\sqrt{2}}|\Phi_3\rangle\langle\Phi_2|\\
&\hspace{-1.7cm}V_{L}^{(1)}(\omega_{24})=\frac{1}{\sqrt{2}}|\Phi_4\rangle\langle\Phi_2|,\quad\,\,V_{L}^{(2)}(\omega_{24})=V_{R}^{(2)}(\omega_{24})=\frac{1}{\sqrt{2}}|\Phi_4\rangle\langle\Phi_2|.
\end{align}
\end{widetext}
By substituting Eqs (\ref{E7}) and (\ref{E8}) into the equation of motion (\ref{E6}), we can obtain the master equation.

\begin{eqnarray}
\label{Dis}
\dot{\rho}(t)=-i[H_s,\rho(t)]+\mathcal{D}_{L}(\rho)+\mathcal{D}_{R}(\rho),
\end{eqnarray}

where  $\mathcal{D}_{\nu}(\rho)$ is the dissipator associated with bath $\mathcal{R}_{\nu}$

\begin{widetext}
\begin{align}
\label{E9}
&\mathcal{D}_{\nu}(\rho)=\gamma_{13}^{(\nu,e)}
L_{\tau_{31}}+\gamma_{13}^{(\nu,a)}L_{\tau_{13}}+\gamma_{14}^{(\nu,e)}
L_{\tau_{41}}+\gamma_{14}^{(\nu,a)}L_{\tau_{14}}
+\gamma_{23}^{(\nu,e)}
L_{\tau_{32}}+\gamma_{23}^{(\nu,a)}L_{\tau_{23}}+\gamma_{24}^{(\nu,e)}
L_{\tau_{42}}+\gamma_{24}^{(\nu,a)}L_{\tau_{24}},
\end{align}
\end{widetext}

where $\tau_{ij}=|\Phi_i\rangle\langle\Phi_j|$, $L_{X}=X\rho X^{\dagger}-\frac{1}{2}\{\rho,X^{\dagger}X\}$ and

\begin{widetext}
\begin{align}
\label{E10-a}
&&\hspace{-3cm}\gamma_{13}^{(L,e)}=
\frac{\kappa\omega_{13}\left(\epsilon-1\right)^2}{2}\left[1+n_{L}\left(\omega_{13}\right)\right],\quad\quad
\gamma_{13}^{(R,e)}=
\frac{\kappa\omega_{13}}{2}\left[1+n_{R}\left(\omega_{13}\right)\right],\\
&&\hspace{-3cm}\gamma_{13}^{(L,a)}=
\frac{\kappa\omega_{13}\left(\epsilon-1\right)^2}{2}\,n_{L}\left(\omega_{13}\right),\quad\quad\quad\quad\,\,
\gamma_{13}^{(R,a)}=
\frac{\kappa\omega_{13}}{2}\,n_{R}\left(\omega_{13}\right),\\
&&\hspace{-3cm}\gamma_{14}^{(L,e)}=
\frac{\kappa\omega_{14}\left(\epsilon+1\right)^2}{2}\left[1+n_{L}\left(\omega_{14}\right)\right],\quad\quad
\gamma_{14}^{(R,e)}=
\frac{\kappa\omega_{14}}{2}\left[1+n_{R}\left(\omega_{14}\right)\right],\\
&&\hspace{-3cm}\gamma_{14}^{(L,a)}=
\frac{\kappa\omega_{14}\left(\epsilon+1\right)^2}{2}\,n_{L}\left(\omega_{14}\right),\,\,\quad\quad\quad\quad
\gamma_{14}^{(R,a)}=
\frac{\kappa\omega_{14}}{2}\,n_{R}\left(\omega_{14}\right),\\
&&\hspace{-3cm}\gamma_{23}^{(L,e)}=
\frac{\kappa\omega_{23}\left(\epsilon-1\right)^2}{2}\left[1+n_{L}\left(\omega_{23}\right)\right],\quad\quad
\gamma_{23}^{(R,e)}=
\frac{\kappa\omega_{23}}{2}\left[1+n_{R}\left(\omega_{23}\right)\right],\\
&&\hspace{-3cm}\gamma_{23}^{(L,a)}=
\frac{\kappa\omega_{23}\left(\epsilon-1\right)^2}{2}\,n_{L}\left(\omega_{23}\right),\quad\quad\,\,\quad\quad
\gamma_{23}^{(R,a)}=
\frac{\kappa\omega_{23}}{2}\,n_{R}\left(\omega_{23}\right),\\
&&\hspace{-3cm}\gamma_{24}^{(L,e)}=
\frac{\kappa\omega_{24}\left(\epsilon+1\right)^2}{2}\left[1+n_{L}\left(\omega_{24}\right)\right],\quad\quad
\gamma_{24}^{(R,e)}=
\frac{\kappa\omega_{24}}{2}\left[1+n_{R}\left(\omega_{24}\right)\right],\\
&&\hspace{-3cm}\gamma_{24}^{(L,a)}=
\frac{\kappa\omega_{24}\left(\epsilon+1\right)^2}{2}\,n_{L}\left(\omega_{24}\right),\quad\quad\,\,\quad\quad
\gamma_{24}^{(R,a)}=\frac{\kappa\omega_{24}}{2}\,n_{R}\left(\omega_{24}\right),
\label{E10-h}
\end{align}
\end{widetext}

In the above equations, $\gamma_{ij}^{(\nu,e)}$ ($\gamma_{ij}^{(\nu,a)}$) represents the emission (absorption) rate from $|\Phi_i\rangle$ to $|\Phi_j\rangle$ ($|\Phi_j\rangle$ to $|\Phi_i\rangle$) due to the interaction of the system with reservoir $\nu$. According to above equations, the emission and absorption rates $\gamma_{13}^{(L,e)}$, $\gamma_{13}^{(L,a)}$, $\gamma_{23}^{(L,e)}$, and $\gamma_{23}^{(L,a)}$  clearly are zero in the asymmetric coupling case ($\epsilon=1$). This implies that the energy transitions $|\Phi_1\rangle\leftrightarrow|\Phi_3\rangle$ and $|\Phi_2\rangle\leftrightarrow|\Phi_3\rangle$, induced by the left reservoir, are not possible, as depicted in Fig.\ref{transitions} pabel (c).\\
The steady-state solution associated with the dynamics governed by the master Eq.\ref{E9} can be obtained by solving the equation $d\rho/dt=0$. It is important to note that the solution of Eq.\ref{E9} consists of two parts: the population components, represented by $P_{1}$, $P_{2}$, $P_{3}$, and $P_{4}$, and the coherence part, which includes the off-diagonal elements $P_{ij}$. In the eigenbasis of $H_S$ \cite{Breuer}, the off-diagonal elements are decoupled from the diagonal elements, resulting in their vanishing in the steady-state limit. Consequently, in the steady state regime, the density matrix in the eigenbasis of $H_S$ becomes a diagonal matrix with the following elements:
\begin{eqnarray}
&&P_{1}=\frac{E_{13}r_1+E_{14}r_2}{r_1(A_{13}+A_{14})}P_{3},\\
&&P_{2}=\frac{A_{23}r_1+A_{24}r_2}{r_1(E_{23}+E_{24})}P_{3},\\
&&P_{3}=\left[\frac{E_{13}r_1+E_{14}r_2}{r_1(A_{13}+A_{14})}+\frac{A_{23}r_1+A_{24}r_2}
{r_1(E_{23}+E_{24})}+\frac{r_2}{r_1}+1\right]^{-1},\\
&&P_{4}=\frac{r_2}{r_1}P_{3},
\end{eqnarray}
with
\begin{widetext}
\begin{align}
&&\hspace{-1.5cm}r_1=\frac{A_{13}E_{14}}{(A_{23}+E_{13})(A_{13}+A_{14})}+\frac{E_{23}A_{24}}{(A_{23}+E_{13})(E_{23}+E_{24})}\nonumber\\
&&\hspace{-1.5cm}r_2=1-\frac{A_{13}E_{13}}{(A_{23}+E_{13})(A_{13}+A_{14})}-\frac{E_{23}A_{23}}{(A_{23}+E_{13})(E_{23}+E_{24})}\nonumber\\
&&\hspace{-1.5cm}A_{ij}=\gamma_{ij}^{(L,a)}+\gamma_{ij}^{(R,a)},\quad\quad
E_{ij}=\gamma_{ij}^{(L,e)}+\gamma_{ij}^{(R,e)}.
\end{align}
\end{widetext}

\section{\label{AppB} $P_{i}^{h}$ at the end of the 3-4 processes in asymmetric GQOC}
As discussed in Appendix (\ref{AppA}), the asymmetric configuration ($\epsilon=1$) results in zero values for certain emission and absorption rates, namely $\gamma_{13}^{(L,e)}$, $\gamma_{13}^{(L,a)}$, $\gamma_{23}^{(L,e)}$ and $\gamma_{23}^{(L,a)}$. Furthermore, during non-equilibrium \emph{processes (3-4)} and in the high temperature gradient limit, the sum of emission and absorption rates $E_{ij}$ and $A_{ij}$, respectively can be simplified as follows:

\begin{align}
A_{ij}&=\gamma_{ij}^{(R,a)},\\
E_{13}&=\gamma_{13}^{(R,e)},\\
E_{23}&=\gamma_{23}^{(R,e)},\\
E_{14}&=\gamma_{14}^{(R,e)}+\frac{\kappa \omega_{14}}{2},\\
E_{24}&=\gamma_{24}^{(R,e)}+\frac{\kappa \omega_{24}}{2}.
\label{gammas-simplyfied}
\end{align}

Since the only dependence on the reservoir temperatures appears in the terms $\gamma_{ij}^{(R,a/e)}$ on the right side of the above equations, it can be concluded that $E_{ij}$ and $A_{ij}$ are independent of the temperature of the left (cold) reservoir. Therefore, only the right (hot) reservoir affects the system and is capable of thermalizing it.  It is evident that the main distinction between this non-equilibrium scenario and the one where only the hot right reservoir is coupled is the addition of $\frac{\kappa \omega_{ij}}{2}$ to the last two equations. This indicates that the total emission rates $E_{14}$ and $E_{24}$ now include a term with a linear dependence on the energy differences $\omega_{ij}$ yet remain independent of the left reservoir. Consequently, we can conclude that the right (hot) reservoir exerts a significant impact on the system, causing the population distribution to converge to a maximally mixed state in high-temperature regimes.

\nocite{*}

\bibliography{apssamp}

\end{document}